\documentclass{PoS}

\usepackage{xspace}

\newcommand{\et}{E_{\rm T}\xspace} 
\newcommand{\met}{$\not\!\!\et$\xspace} 
\newcommand{\pt}{\ensuremath{p_{\rm T}}\xspace} 
\newcommand{\ttbar}{\ensuremath{\rm t\bar{t}}\xspace} 
\newcommand{\tprime}{\ensuremath{\rm t'}\xspace}

\newcommand{\tprimetobW}{\tprime \ensuremath{\to} b W\xspace}
\newcommand{\tprimetotZ}{\tprime \ensuremath{\to} t Z\xspace}
\newcommand{\tprimetotH}{\tprime \ensuremath{\to} t H\xspace}
\newcommand{\bprime}{\ensuremath{\rm b'}\xspace}

\newcommand{\bprimetotW}{\bprime \ensuremath{\to} t W\xspace}
\newcommand{\bprimetobZ}{\bprime \ensuremath{\to} b Z\xspace}

\newcommand{\tfivethree}{\ensuremath{\rm T_{5/3}}\xspace} 
\newcommand{\tfivethreetotW}{\tfivethree \ensuremath{\to} t W\xspace}

\title{Search for exotic heavy top and bottom quark partners with CMS}

\ShortTitle{Heavy quark search in CMS}

\author{\speaker{Devdatta Majumder}\\
        National Taiwan University,\\
        On behalf of the CMS Collaboration \\ 
        E-mail: \email{devdatta.majumder@cern.ch}}

\abstract{We present searches for massive top and bottom quark partners at CMS using LHC pp collision data collected at centre-of-mass energy $\sqrt{s} = 8$~TeV. Such partners can be found in models predicting vector-like quarks to solve the hierarchy problem and stabilize the Higgs boson mass. The searches span a range of final states containing several lepton and jet multiplicities, and limits were set on mass and production cross sections as a function of branching ratios.} 

\FullConference{The European Physical Society Conference on High Energy Physics -EPS-HEP2013\\
		18-24 July 2013\\
		Stockholm, Sweden}

\graphicspath{{./figs/}} 

\begin{document}

\section{Motivation and overview}  


Vector-like quarks, whose left- and right-handed components transform identically under the standard model (SM) electroweak gauge transformation, are predicted in many beyond standard model scenarios to elucidate the nature of electroweak symmetry breaking~\cite{VLQRef}. 
Searches of these vector-like quarks are presented with data collected by the CMS experiment~\cite{CMSRef} from pp collisions at a centre of mass energy $\sqrt{s} = 8$~TeV at the LHC. While no indication of vector-like quarks were found, limits were set on their production cross sections and masses. 
The analyses were performed in a model-independent way. 

\section{Analysis techniques}

\begin{figure}[h]
\begin{center}
\begin{tabular} {cc}
\includegraphics[width=0.29\textwidth]{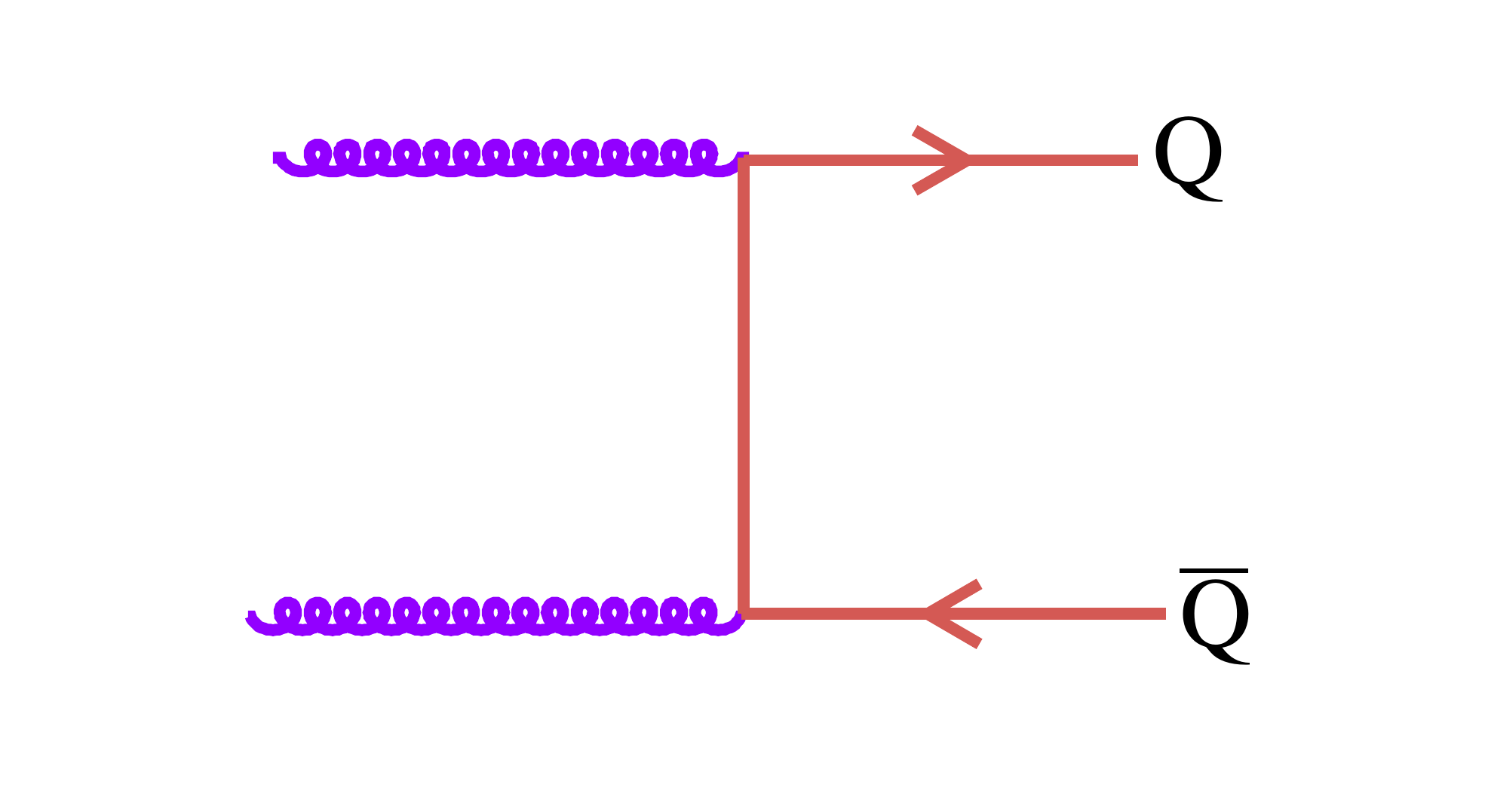} &
\includegraphics[width=0.29\textwidth]{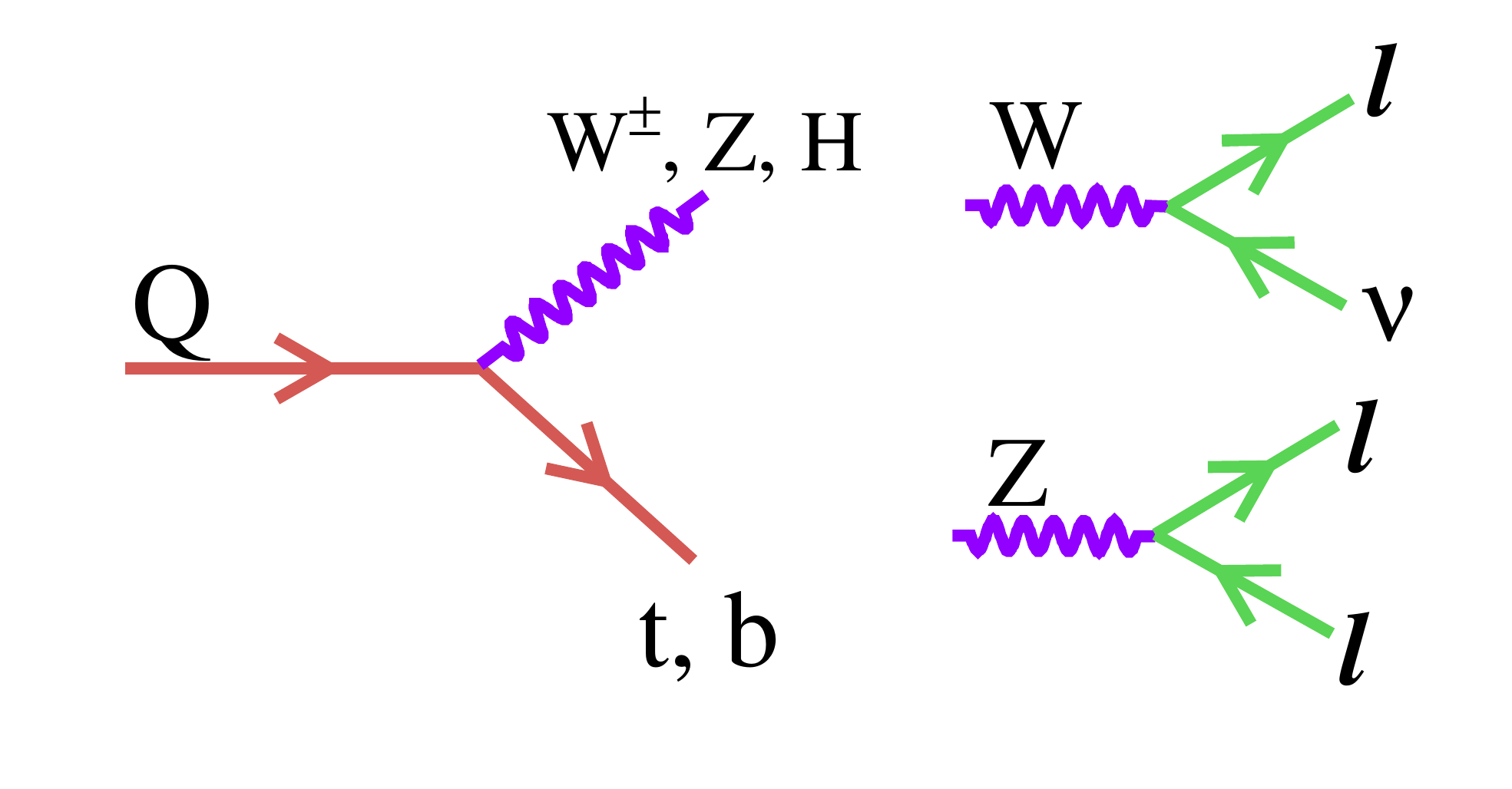} \\
(a) & (b) \\ 
\end{tabular}
\caption{The pair-production of vector-like quarks (a) and their decay channels (b).
} 
\label{fig:VLQ_PairProd_Decay}
\end{center} 
\end{figure}

Pair-production of vector-like quarks, shown in  Fig.~\ref{fig:VLQ_PairProd_Decay} (a), dominates for masses below 1~TeV. 
The possible decay channels, shown in Fig.~\ref{fig:VLQ_PairProd_Decay} (b), are to a W$^{\pm}$, Z or Higgs boson and third-generation quarks (top or bottom). 
Final states with leptons (electrons and muons) were considered, as they provide a relatively clean signature and helps in triggering such events. 

Events were classified according to the lepton multiplicity, flavours and their electric charges: single lepton, oppositely-charged and same-flavoured (OSSF), same charge and flavour (SS2L), oppositely-charged and unlike-flavoured (OSOF). 
Several jet algorithms were used for identifying hadronic jets. 
Jets reconstructed using the anti-$k_{\rm T}$ algorithm with a distance parameter 0.5 (AK5 jets) were used for tagging jets from b quark hadronisation (b jets). 
The Cambridge-Aachen algorithm with a distance parameter of 0.8 (CA8 jets) was used to reconstruct boosted W and top jets. The W- and top-tagging algorithms, based on jet substructure information of the CA8 jets, were used to identify boosted W and top jets. 
The missing transverse energy \met was defined as the negative of sum of the transverse momenta of all reconstructed particles. 
$H_{\rm T}$ was defined as the scaler sum the transverse momenta of all reconstructed jets and $S_{\rm T}$, as the scaler sum of the transverse momenta of all leptons, jets and \met in the event. 


The main backgrounds are the inclusive W and Z boson or \ttbar production in association with jets with lesser contributions from single top quark and diboson production. Rare SM processes like \ttbar+V or \ttbar+VV contribute mainly in final states with SS2L lepton pairs. The signal, the inclusive W and Z boson, \ttbar, and the rare SM processes were simulated using a combination of \textsc{Madgraph} and \textsc{Pythia6} event generator programs. The single top quark process was simulated using \textsc{Powheg}. Signal cross sections were obtained using an approximate next-to-next-to-leading order calculation using the \textsc{Hathor} package. 

The main systematic uncertainties were those on the integrated luminosity ($\pm4.4\%$), the reconstruction and simulation, b-tagging, W- and top-tagging. The uncertainties on the simulated samples were due to the parton distribution functions, and theoretical uncertainties. The jet energy corrections and background estimation uncertainties were also taken into account. 

\section{Search for a vector-like top quark \tprime}

\begin{figure}[!htb]
\begin{center}
\includegraphics[width=0.29\textwidth]{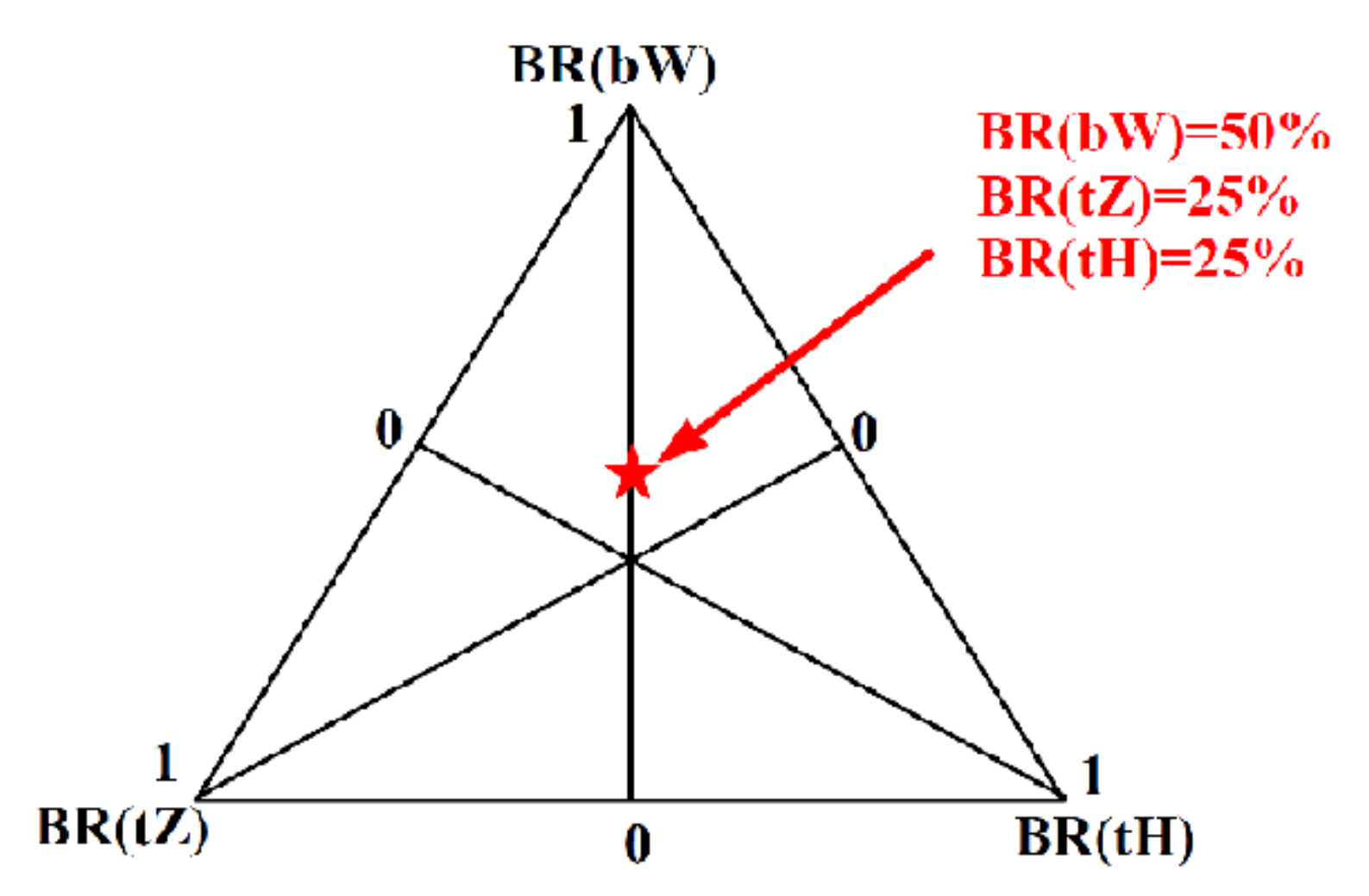} 
\caption{The \tprime quark branching ratio plane over which mass limits were set. A special combination of branching ratios is marked by a star, which corresponds to a model where the \tprime quark decays democratically via the charged-current (\tprimetobW) and the neutral current (\tprimetotZ or \tprimetotH) processes~\cite{TprimeDecayRef}.}  
\label{fig:TrianglePlot}
\end{center} 
\end{figure}

The decay channels explored were \tprime$\to$bW, \tprime$\to$tZ and \tprime$\to$tH. Two analyses, one with a single lepton  final state, and the other with a multilepton final state, were performed~\cite{CMS-PAS-B2G-12-015}.  The limit on the mass of the \tprime quark was set for all allowed sets of branching ratios, as depicted in Fig.~\ref{fig:TrianglePlot}. 

\begin{figure}[!htb]
\begin{center}
\begin{tabular} {ccc}
\includegraphics[width=0.29\textwidth]{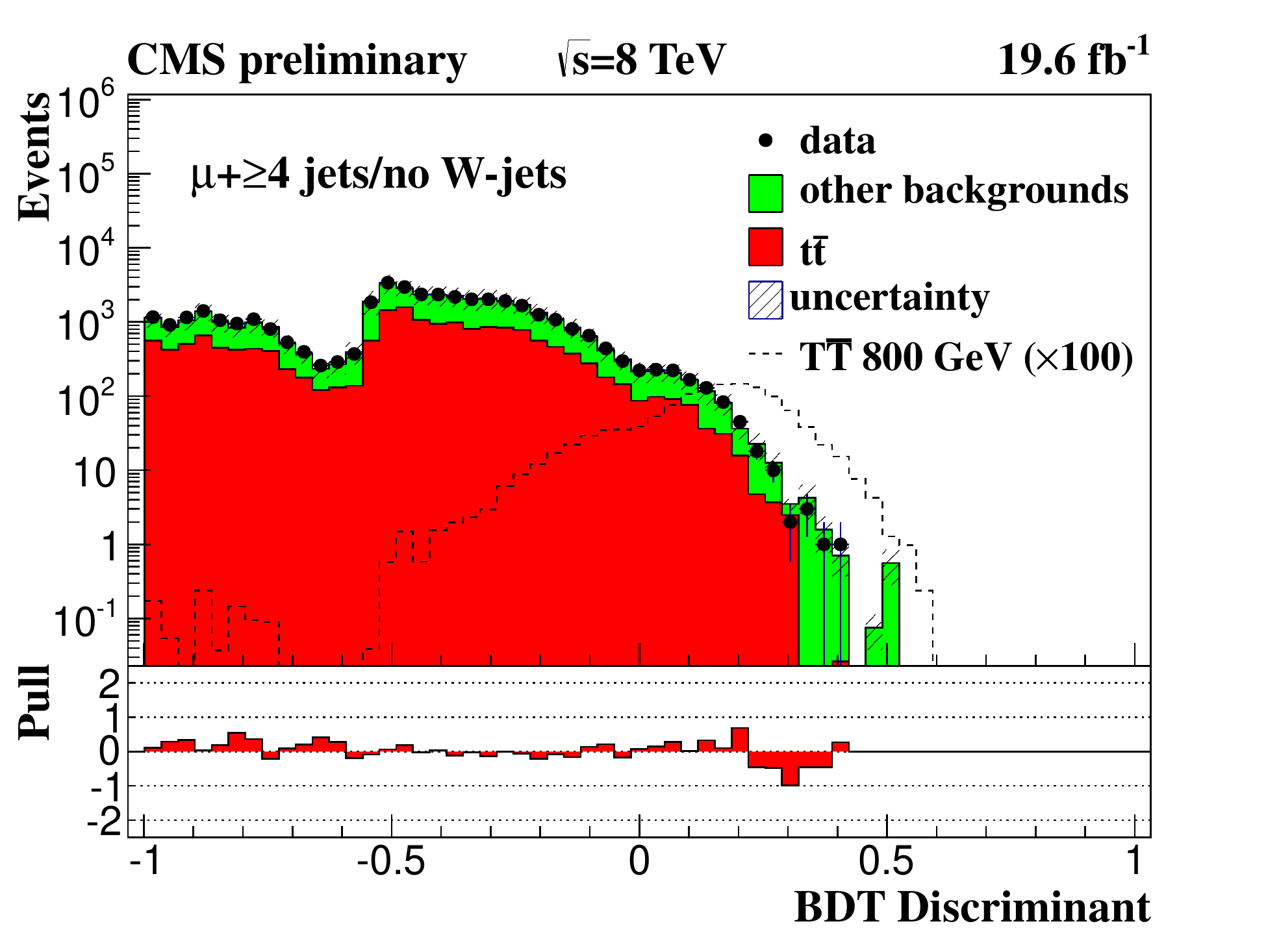} & 
\includegraphics[width=0.29\textwidth]{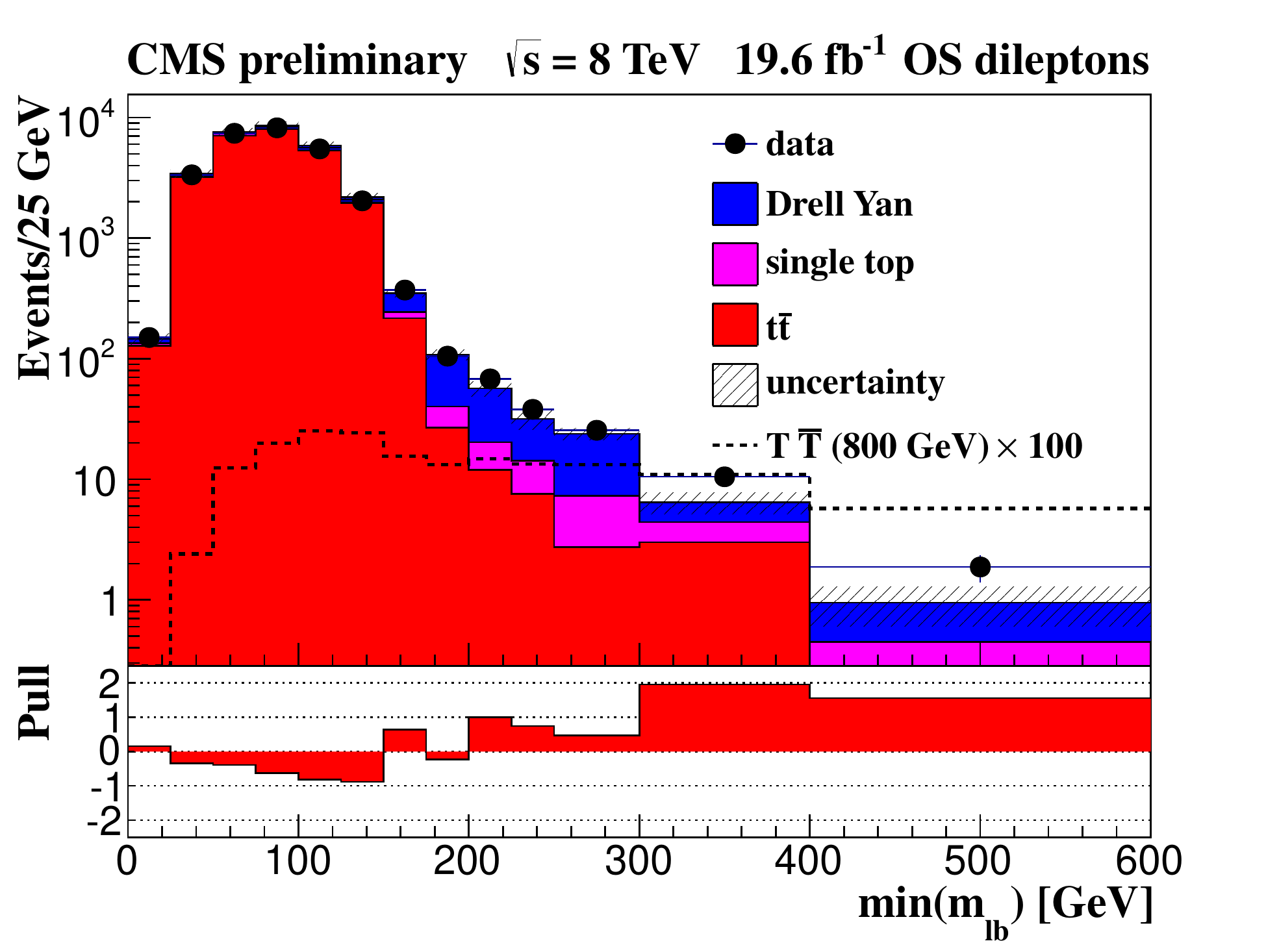} & 
\includegraphics[width=0.29\textwidth]{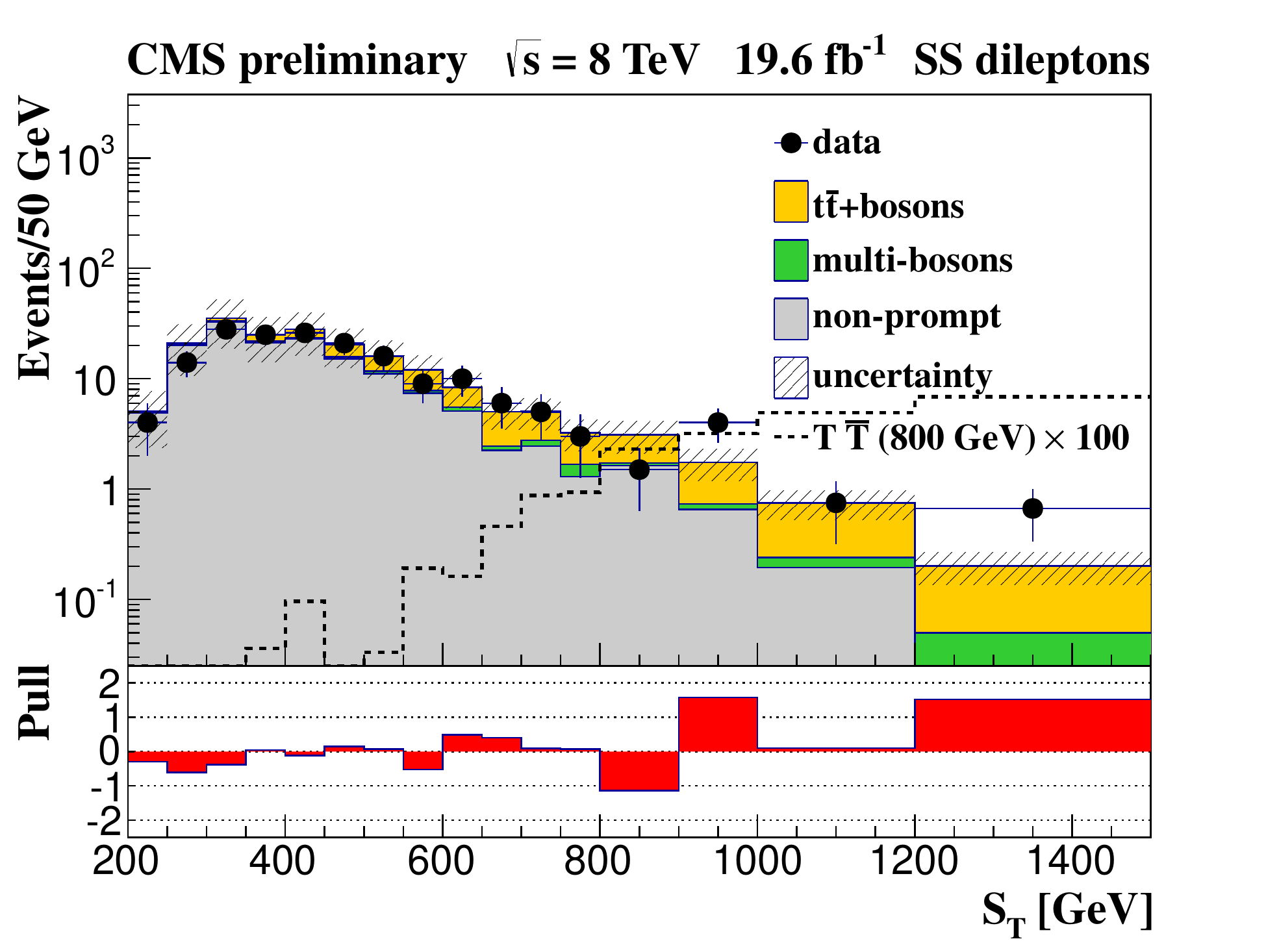} \\ 
(a) & (b) & (c)\\ 
\end{tabular}
\caption{The BDT distribution for the single lepton final state (a), the min($M(\ell {\rm b})$) distribution in OS23 events (b) and the $S_{\rm T}$ distribution (c) in SS2L final states in the search for the \tprime quark.}  
\label{fig:Tprime_1LBDT_OS2L_SS2L}
\end{center} 
\end{figure}

The single lepton analysis required one lepton with \pt $> 32$~GeV and $|\eta| < 2.1$. Events were selected and categorized according to the number of b-tagged and W-tagged jets in the final state. 
The signal and background separation was obtained by training a boosted decision tree (BDT) separately for events with an electron or a muon, and in two categories, based on the presence and absence of a W-tagged jet. 
The background modelling was validated in a data control sample without b-tagged jets. The BDT discriminator shape for one event category is shown in Fig.~\ref{fig:Tprime_1LBDT_OS2L_SS2L} (a).  

The multilepton analysis used three event categories: the opposite-signed (OS) and the same-signed (SS) dilepton, and the trilepton. The OS category further contained OSSF and OSOF type events. 
The number of jets, the $H_{\rm T}$ and the $S_{\rm T}$ were used to reject backgrounds. 
The OS23 event category, where the invariant mass of the two leptons were inconsistent with that from a Z boson (Z-veto), was used to enhance the sensitivity of the \tprime\tprime$\to$bWbW channel. 
The variable min($M(\ell, {\rm b\,jet})$), which is the minimum of the invariant masses of all lepton-b jet pairs, was used to obtain a good rejection of the \ttbar+jets background in such events. 
A total of twelve event categories were used, based on the number of leptons, the lepton flavours and the Z-veto. The  min($M(\ell, {\rm b\,jet})$) for OS23 events, and the $S_{\rm T}$ distributions in SS2L final states are shown in Figs.~\ref{fig:Tprime_1LBDT_OS2L_SS2L} (b) and ~\ref{fig:Tprime_1LBDT_OS2L_SS2L} (c), respectively. 

\begin{figure}[!htb]
\begin{center}
\begin{tabular} {ccc}
\includegraphics[width=0.29\textwidth]{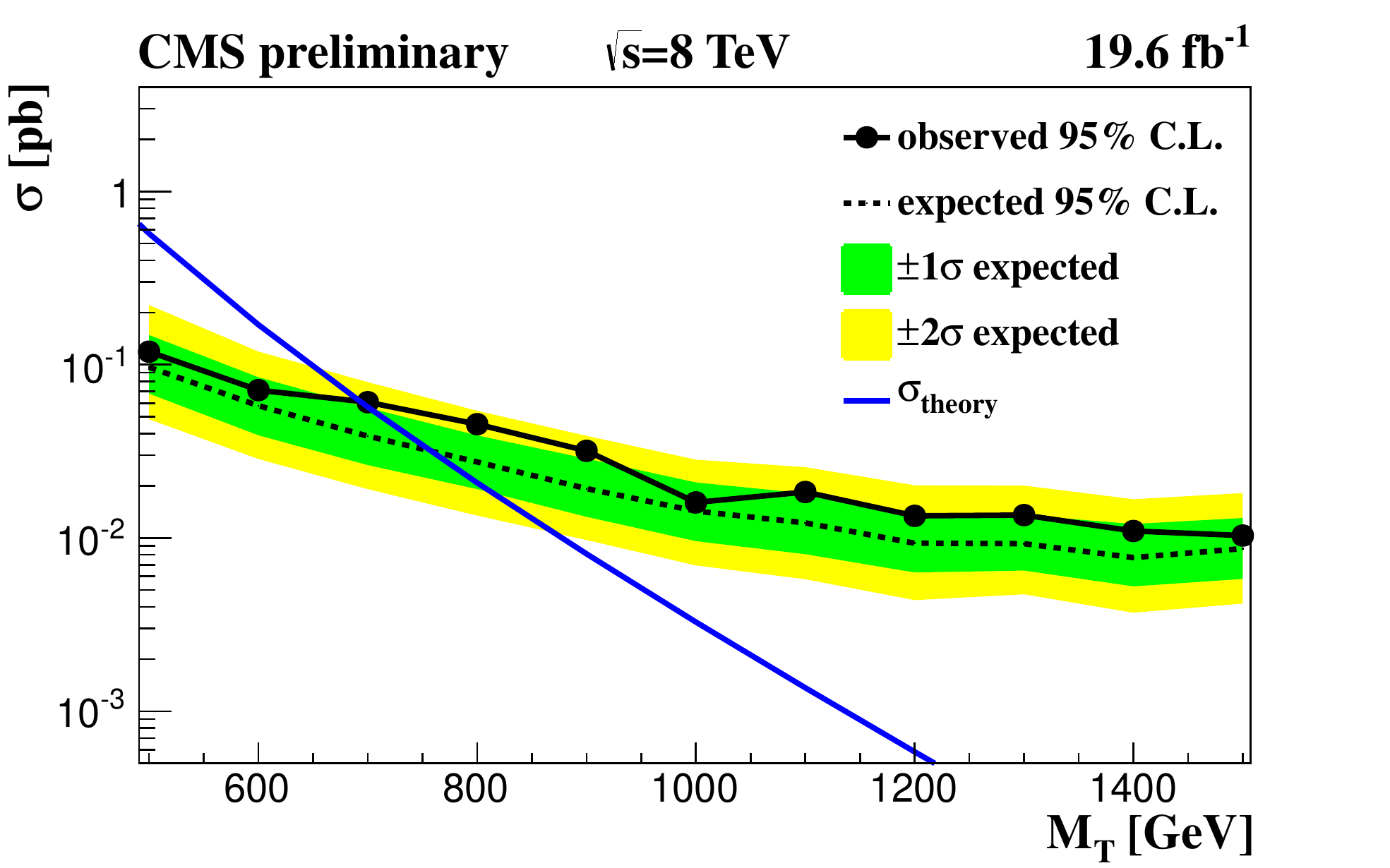} & 
\includegraphics[width=0.29\textwidth]{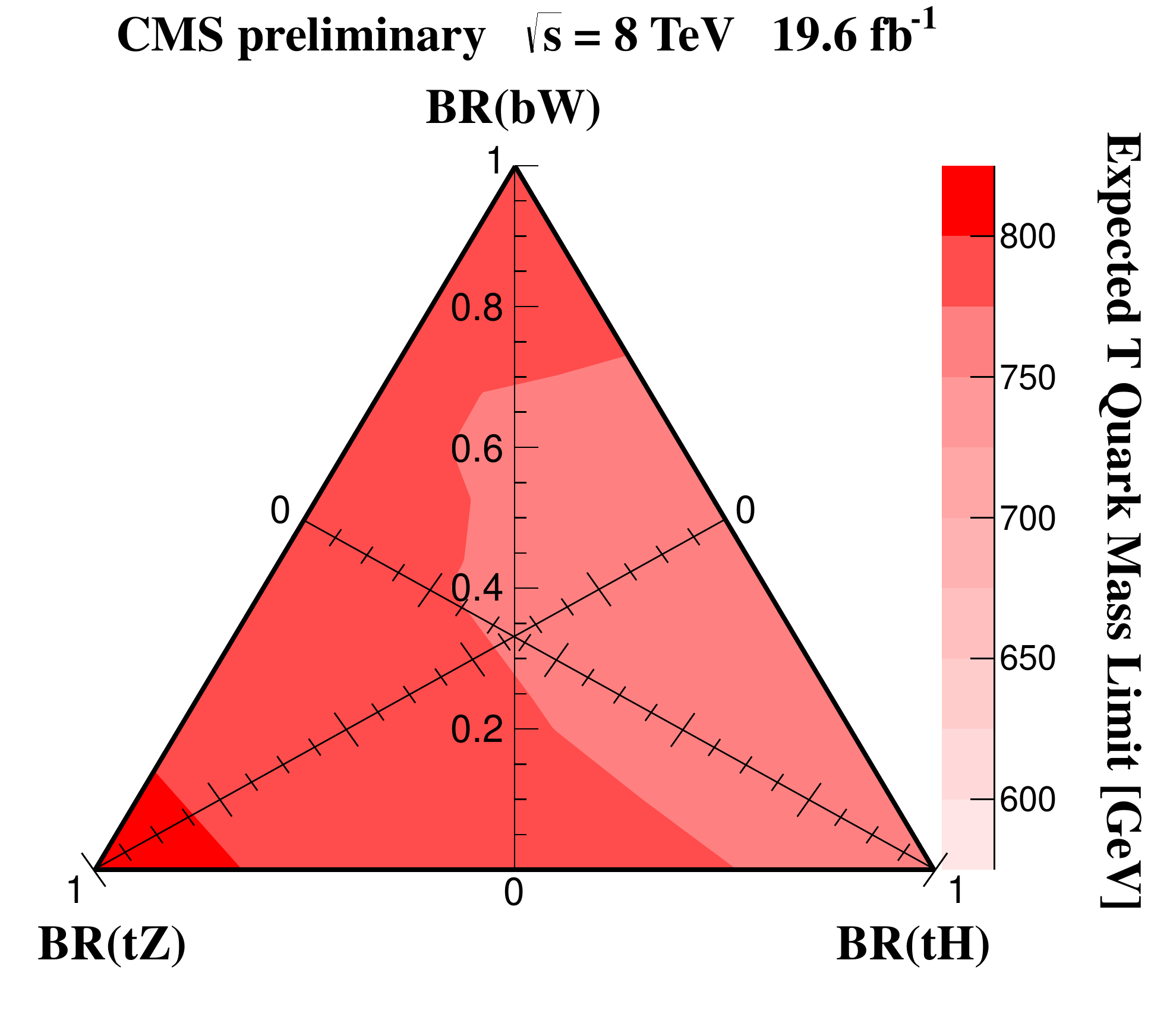} & 
\includegraphics[width=0.29\textwidth]{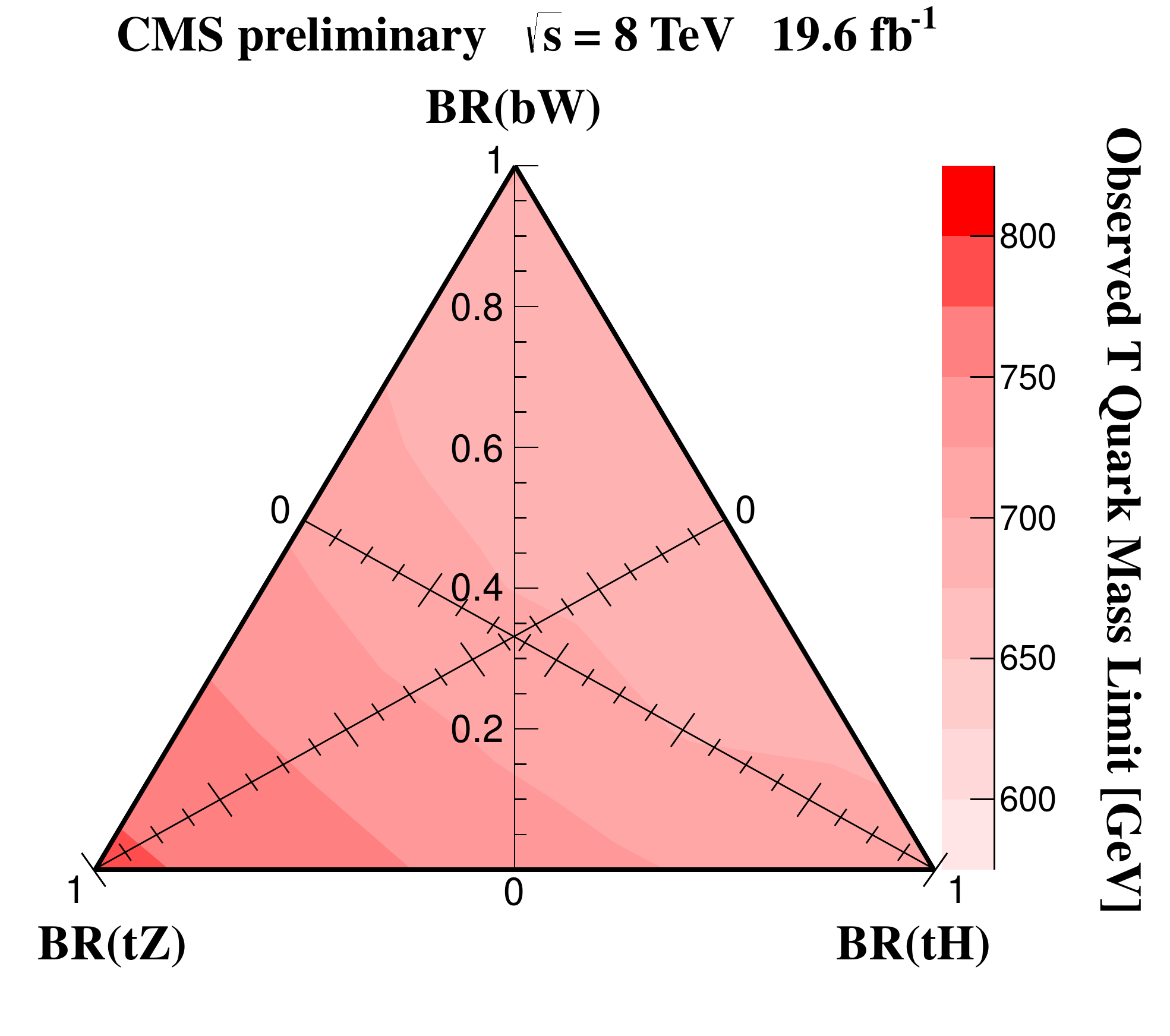} \\ 
(a) & (b) & (c) \\ 
\end{tabular}
\caption{The 95\% upper limit on the \tprime pair-production cross section as a function of $M(\tprime)$ (a). The 95\% expected (b) and observed (c) lower limits on $M(\tprime)$ for different branching fractions of the \tprime.}  
\label{fig:Tprime_CSLimit_MassLimits}
\end{center} 
\end{figure}

The \tprime mass limits were obtained by using a likelihood fit of the BDT shapes in signal and background to the data. The multilepton analysis used counting of signal and background events in the 12 event categories. The observed and the expected lower mass limits for $BR(\tprimetobW) = 50\%$ and $BR(\tprimetotZ) = BR(\tprimetotH) = 25\%$ are  696~GeV and 773~GeV respectively, at 95\% confidence level (CL), as shown in Fig.~\ref{fig:Tprime_CSLimit_MassLimits} (a).
The observed and expected mass limits for the whole range of \tprime branching ratios is shown in Figs.~\ref{fig:Tprime_CSLimit_MassLimits} (b) and ~\ref{fig:Tprime_CSLimit_MassLimits} (c), respectively.

\section{Search for a charge $5{\rm e}/3$ top quark partner \tfivethree}

\begin{figure}[!htb]
\begin{center}
\begin{tabular}{ccc}
\includegraphics[width=0.28\textwidth]{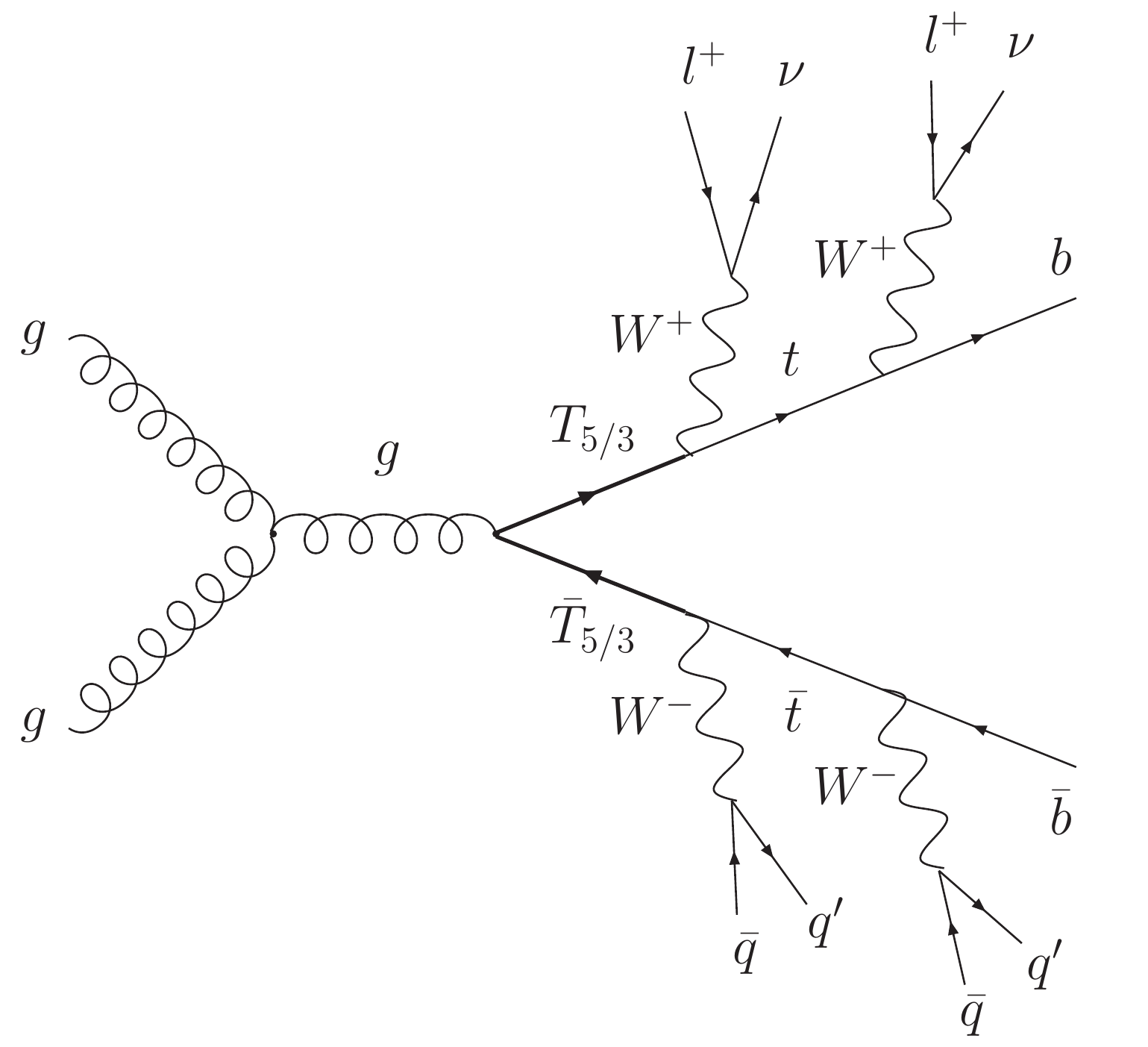} &
\includegraphics[width=0.28\textwidth]{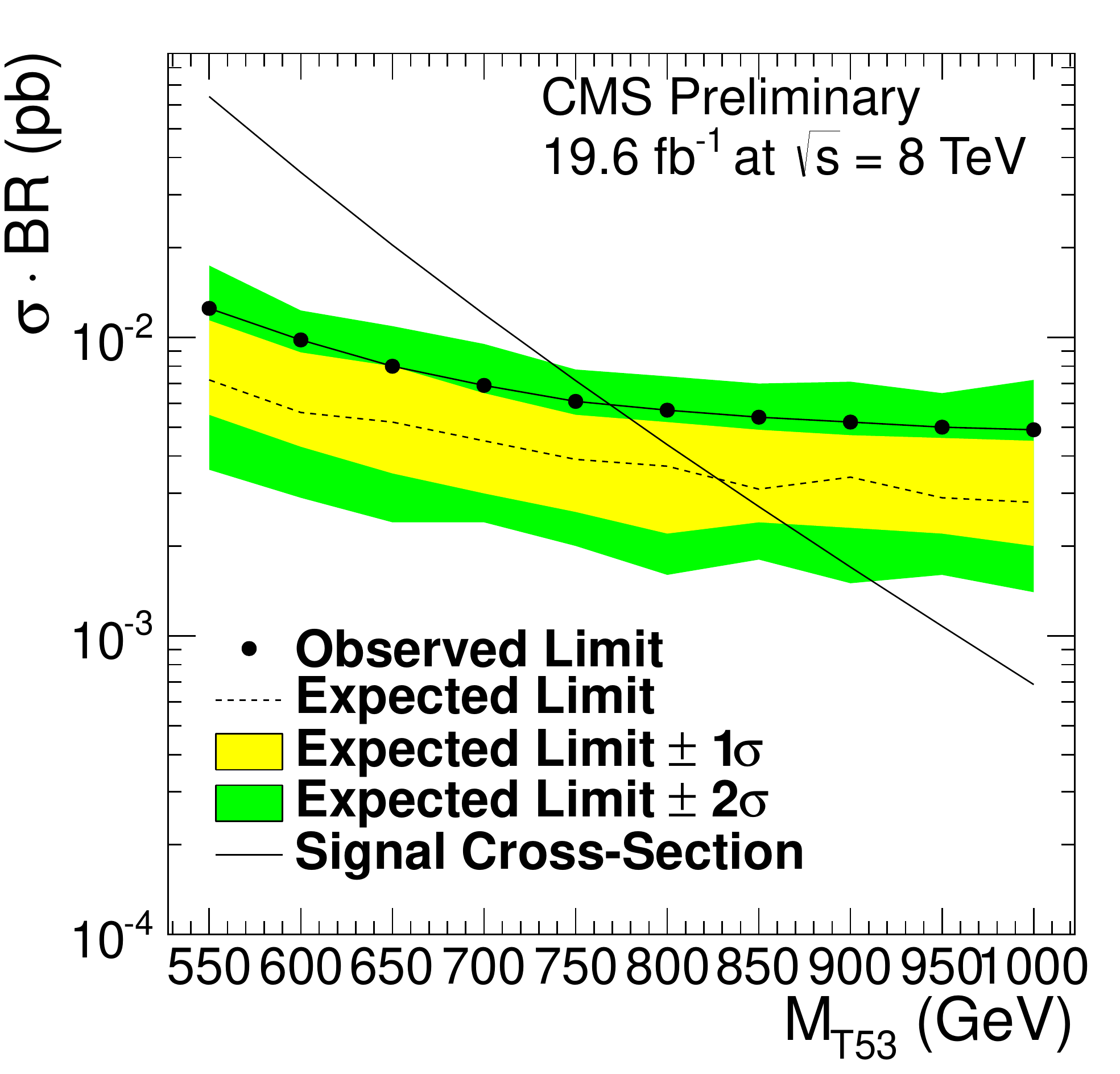} &  
\includegraphics[width=0.28\textwidth]{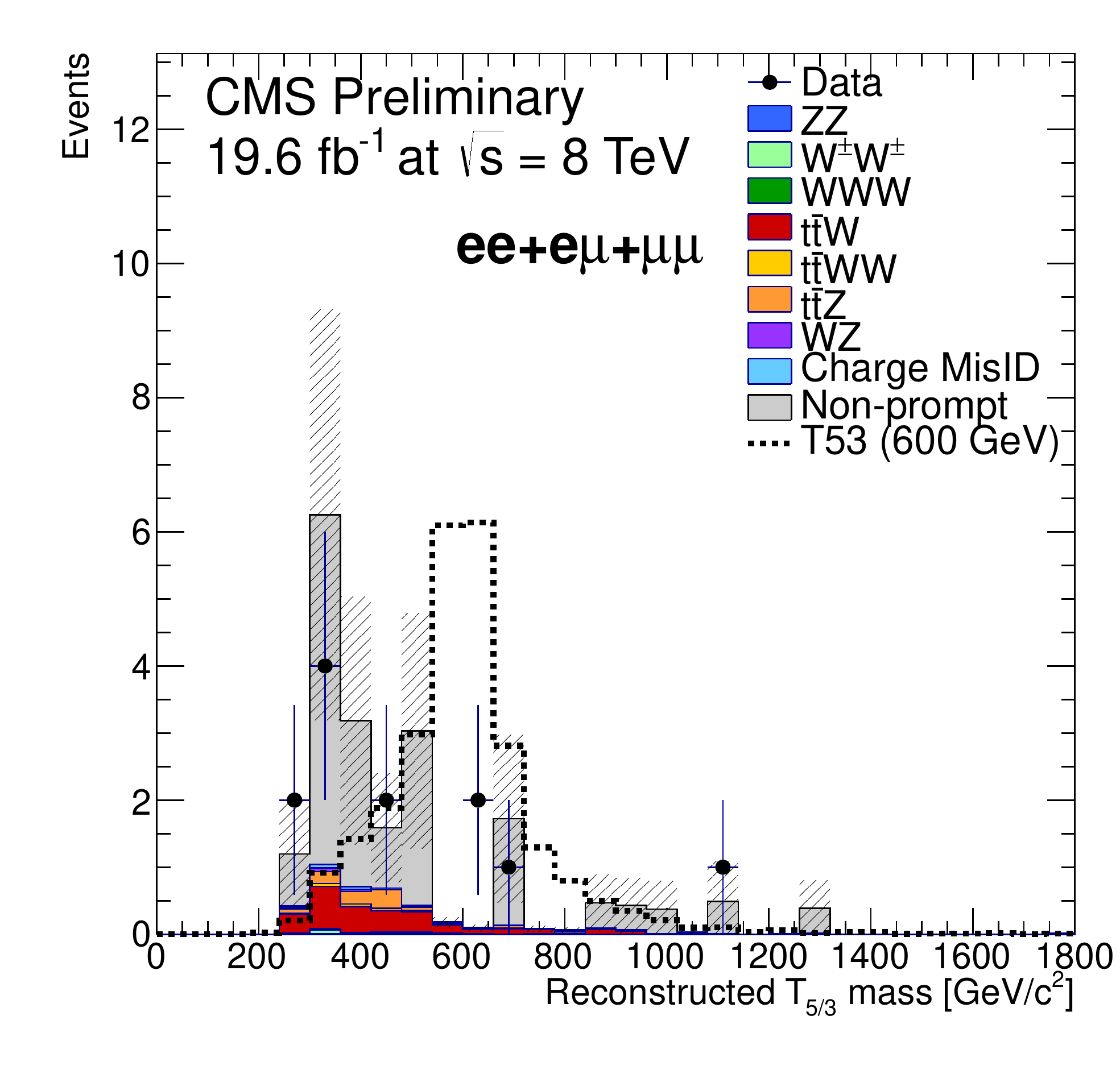} \\ 
(a) & (b) & (c) \\ 
\end{tabular} 
\caption{The pair-production of a \tfivethree quark (a). The 95\% upper limit on the pair-production cross section of the \tfivethree quark (b) and the invariant mass distribution of the \tfivethree quark candidates (c).}   
\label{fig:T53_FeynDiag_CSLimit_Mass} 
\end{center} 
\end{figure}

The search for a exotic quark of charge $5{\rm e}/3$ was performed~\cite{CMS-PAS-B2G-12-012}. 
The pair-production and decay of the \tfivethree quark is shown in Fig.~\ref{fig:T53_FeynDiag_CSLimit_Mass} (a), with the notable feature that for one \tfivethree decaying leptonically and the other hadronically, the final state contains a SS2L pair, resulting in highly suppressed SM backgrounds. Backgrounds from rare SM processes like \ttbar+W, \ttbar+Z, WWW, same electric charge W boson pair-production were evaluated from simulations.  Events with an OS2L pair may be part of the background composition, due to the misidentification of the lepton's electric charge, with contributions also from instrumental backgrounds due to fake leptons. These were estimated from the data.  The estimated background is 6.6$\pm$2 while the number of observed events is 11. 
The mass limit on the \tfivethree was obtained using a frequentist counting experiment and is shown in Fig.~\ref{fig:T53_FeynDiag_CSLimit_Mass} (b), with the observed lower limit at 770~GeV and the expected at 830~GeV, at 95\% CL. 

Mass reconstruction of \tfivethree can possibly improve the limits in data corresponding to higher integrated luminosity and has been tried out in the current dataset with a looser selection criteria, as shown in Fig.~\ref{fig:T53_FeynDiag_CSLimit_Mass} (c). The masses were reconstructed using jet constituents with a set of combinations, as given in Table~\ref{tab:T53_MassReco}. For a fully hadronically decaying \tfivethree quark, the decay products are a W boson and a top quark, which decays to a W boson and a b quark. These were reconstructed using top-, W- and b-tagged jets which were then combined to reconstruct the \tfivethree candidate.

\begin{table}
\begin{center}
\caption{The combinations of top , W and AK5 jets used to reconstruct a hadronically decaying \tfivethree quark.}
\label{tab:T53_MassReco}
\begin{tabular}{c|c|c|c}
\hline 
\hline 
 & Top jet & W jet & AK5 jet \\
\hline 
Combination 1 & 1 & 1 & - \\ 
\hline 
Combination 2 & 1 & - & 2 (W candidate) \\ 
\hline 
Combination 3 & - & 2 & 1 \\ 
\hline 
Combination 4 & -  & 1 & 2 (W candidate) + 1 \\ 
\hline 
Combination 5 & - & - & 2$\times$2 (W candidate) + 1 \\ 
\hline 
\hline 
\end{tabular}
\end{center}
\end{table}

\section{Search for a vector-like bottom quark \bprime}

\begin{figure}[!htb]
\begin{center}
\begin{tabular} {cc}
\includegraphics[width=0.29\textwidth]{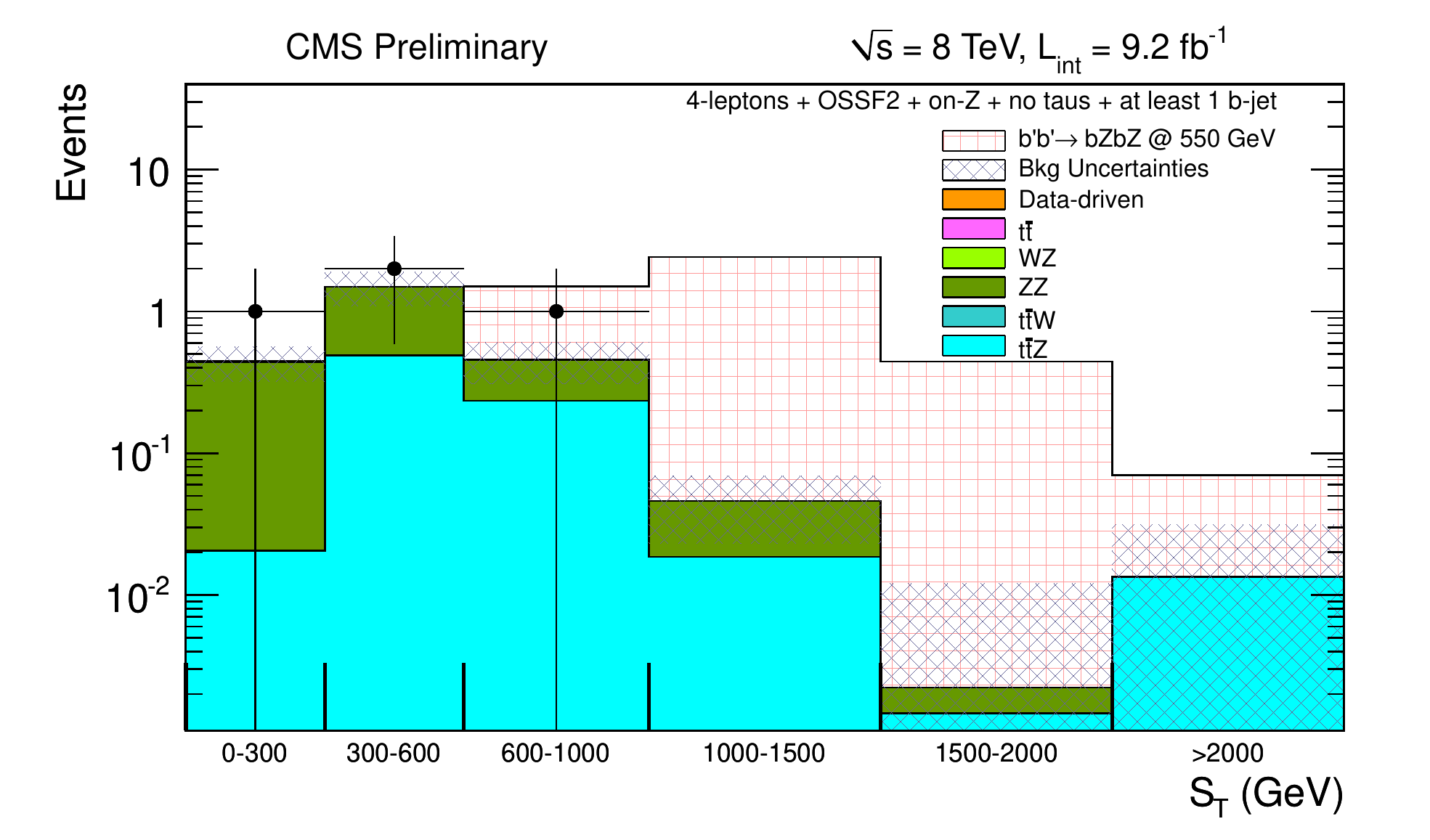} & 
\includegraphics[width=0.29\textwidth]{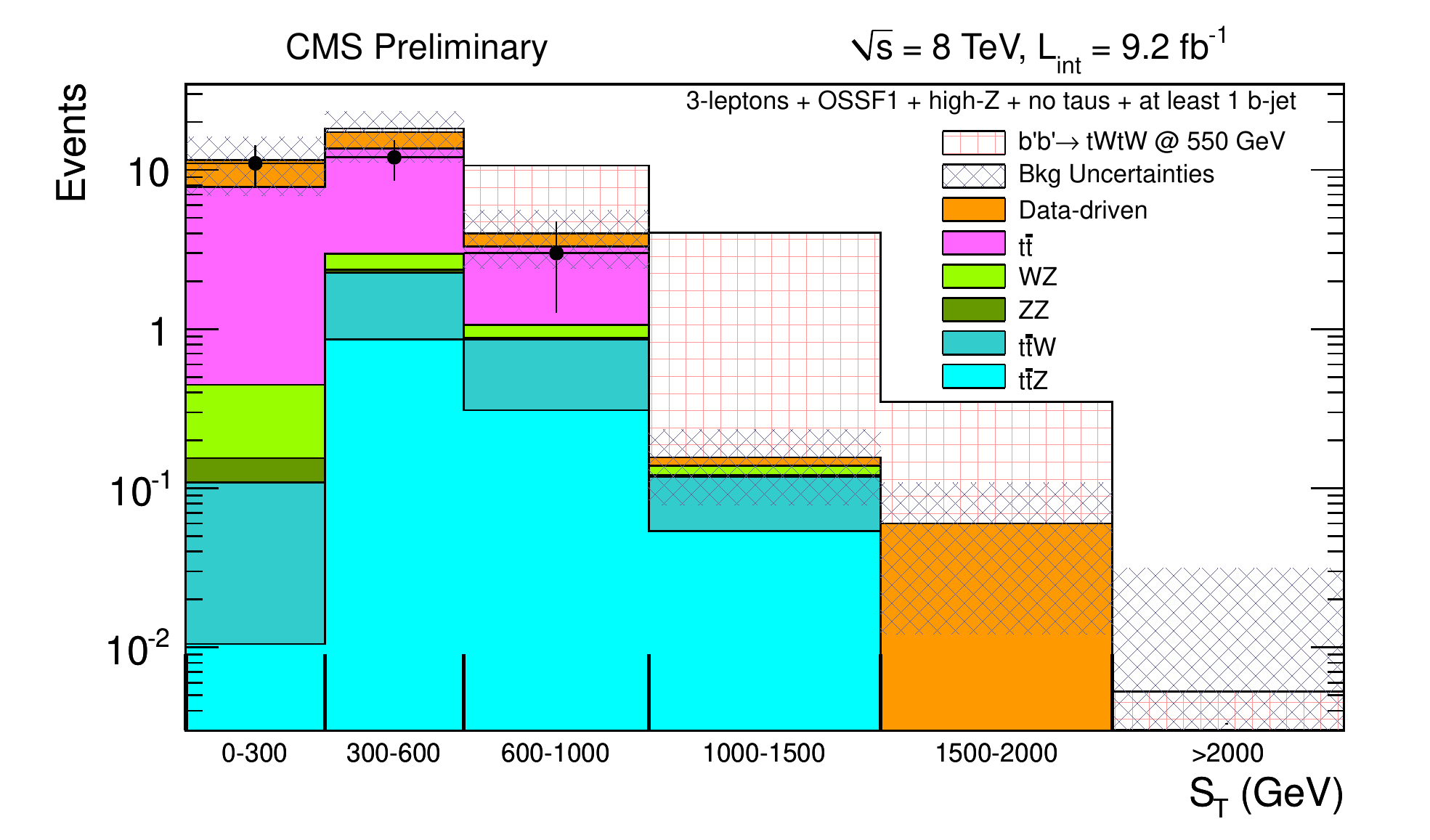} \\ 
(a) & (b) \\ 
\includegraphics[width=0.29\textwidth]{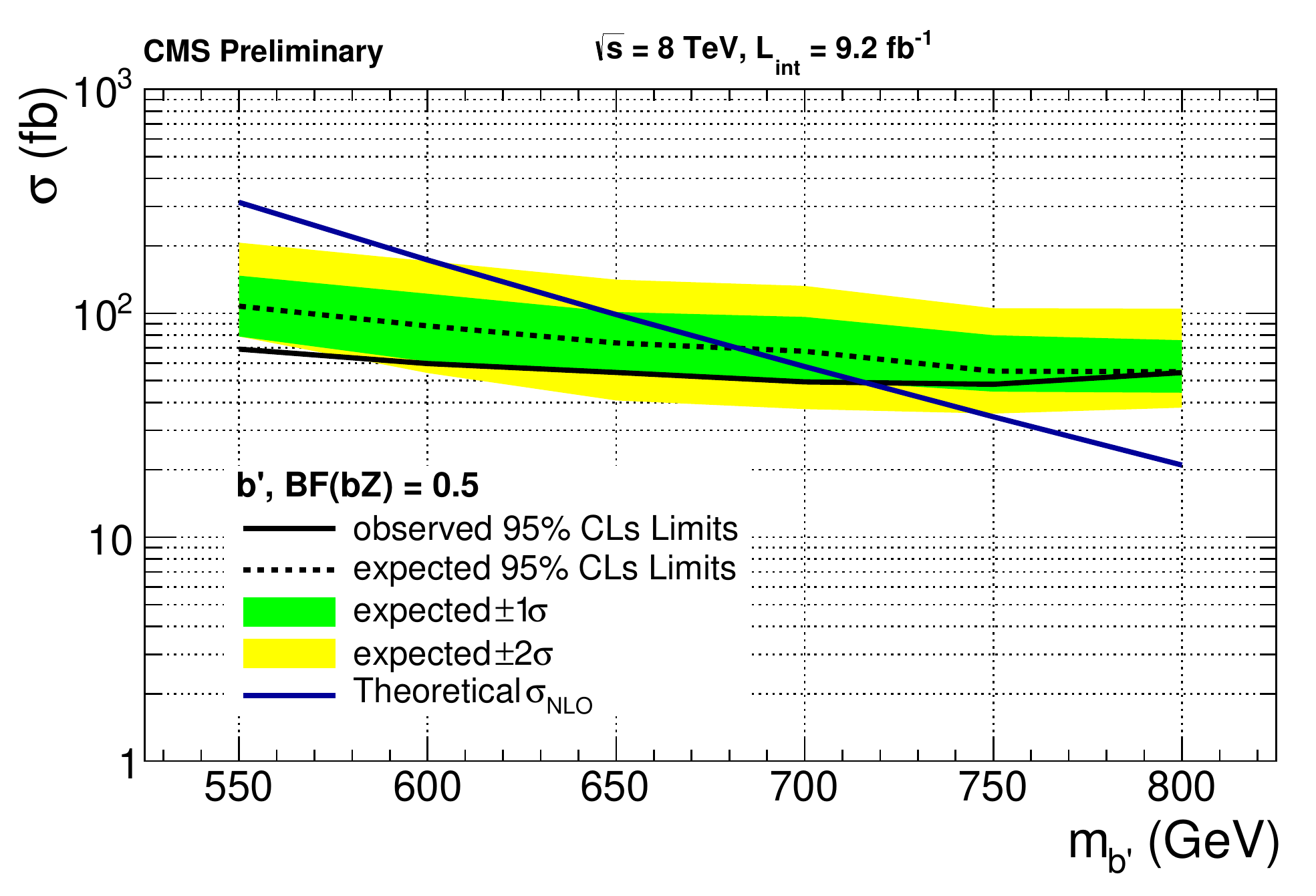} & 
\includegraphics[width=0.29\textwidth]{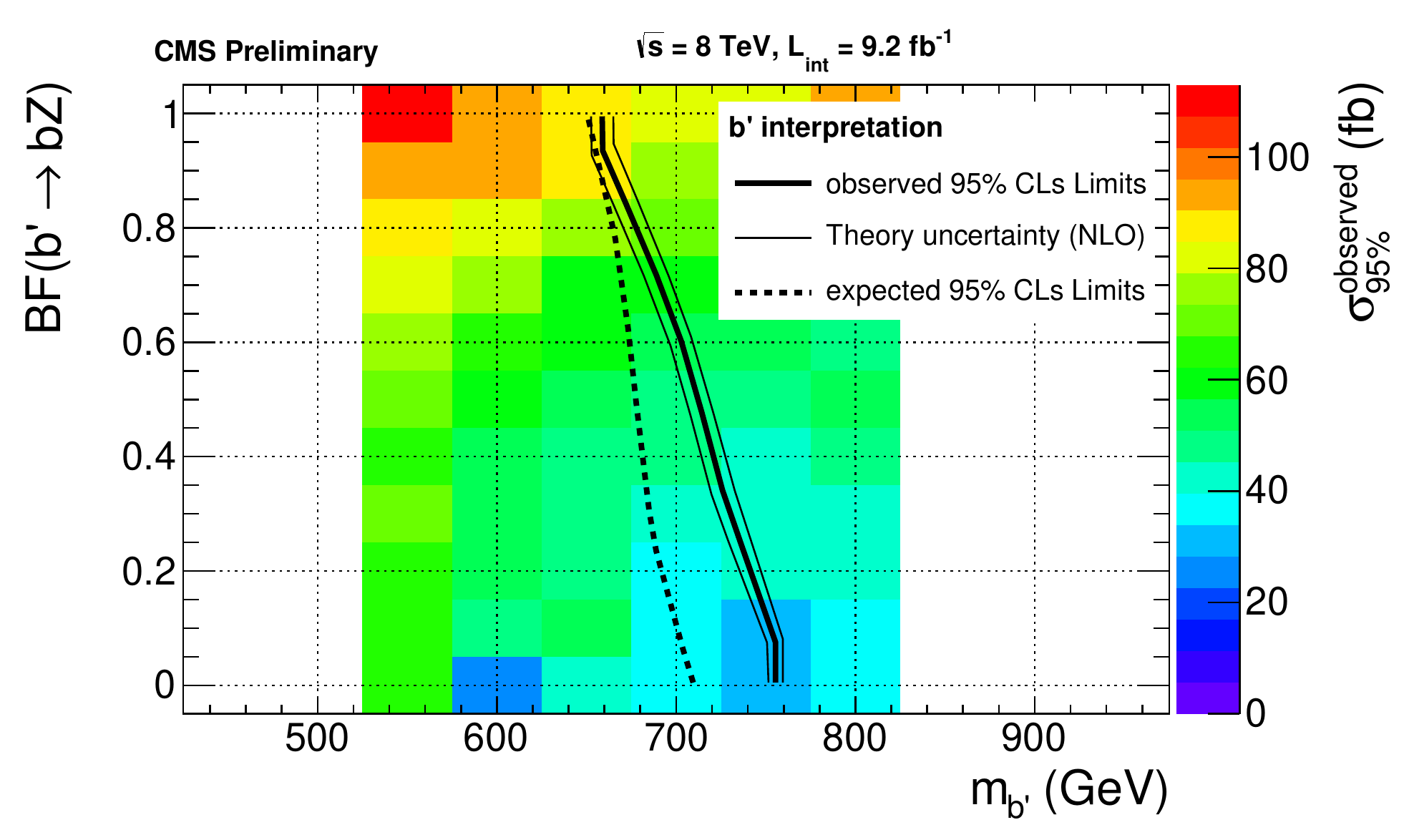} \\ 
(c) & (d) \\ 
\end{tabular}
\caption{$S_{\rm T}$ distribution of events with Z boson + 1 b jet (a) and of events with an opposite-charged same-flavoured lepton pair outside the $M({\rm Z})$ window. The upper limit on the cross section of the \bprime quark pair-production for $BR(\bprimetobZ) = BR(\bprimetotW) = 50\%$ (c) and on the \bprime quark mass as a function of $BR(\bprimetobZ)$ (d) at 95\% CL.}   
\label{fig:Bprime_ST_Limits}
\end{center} 
\end{figure}

A re-interpretation of an $R$-parity supersymmetry search in final states containing three or four leptons and b-tagged jets was used to place limits on the \bprime quark decaying to tW or bZ~\cite{CMS-PAS-SUS-12-027}. Events were categorized as per the lepton flavour, charge and multiplicities, and the number of b-tagged jets. Z boson candidates were reconstructed from OSSF lepton pairs with $75 < M(\ell\ell) < 105$~GeV. The main backgrounds in this search were from misidentified non-prompt leptons, which were estimated using the data. The contributions from \ttbar and dibosons were obtained from simulations and verified in data control samples. The $S_{\rm T}$ distribution of the events in the four lepton+1 b-tagged jets with a Z boson candidate event category is shown in Fig.~\ref{fig:Bprime_ST_Limits} (a) and in the four lepton+1 b-tagged jets with no Z boson candidates event category in Fig.~\ref{fig:Bprime_ST_Limits} (b). \rm 

The level of agreement between the data and the estimated backgrounds was found to be good across all event categories and 
limits were placed on the cross section and mass of the \bprime quark. 
For a \bprime quark decaying to bZ or tW with a branching fraction of 50\% each, the lower limit on the \bprime quark was set at 700~GeV, as shown in Fig.~\ref{fig:Bprime_ST_Limits} (c). The limit on the \bprime quark as function of the $BR(\bprimetobZ)$ is shown in Fig.~\ref{fig:Bprime_ST_Limits} (d) and the \bprime quark in the range of 660-760~GeV is eliminated, depending on the branching ratio. 

\section{Conclusion}

The searches for heavy vector-like quarks in CMS is presented. No evidence for vector-like top or bottom partners was found with the data collected in 2012 pp collisions at $\sqrt{s} = 8$~TeV. The lower limits on the masses of the heavy quarks were set as given in Table~\ref{tab:SummaryOfLimits}. 

\begin{table}[!htb]
\begin{center}
\caption{The summary of the limits obtained for the searches for vector-like quarks presented above, given for a representative set of branching ratios.} 
\label{tab:SummaryOfLimits}
\begin{tabular}{c|c|c} 
\hline 
\hline 
Heavy quark & Branching ratio & Mass limit \\ 
 (electric charge) &  & (GeV) \\ 
\hline
\tprime ($+2/3$)     & $BR$(\tprimetobW):$BR$(\tprimetotZ):$BR$(\tprimetotH) = 0.5:0.25:0.25 & 696 \\ 
\hline
\tfivethree ($+5/3$) & $BR$(\tfivethreetotW) = 1                                             & 770 \\ 
\hline 
\bprime ($-1/3$)     & $BR$(\bprimetotW):$BR$(\bprimetobZ) = 0.5:0.5                         & 700 \\ 
\hline 
\hline 
\end{tabular}
\end{center}
\end{table}

\end{document}